\documentclass[conference]{IEEEtran}
\usepackage{amsmath,amsfonts}
\usepackage{algorithmic}
\usepackage{algorithm}
\usepackage{cite}
\usepackage{array}
\usepackage[caption=false,font=normalsize,labelfont=sf,textfont=sf]{subfig}
\usepackage{textcomp}
\usepackage{stfloats}
\usepackage{url}
\usepackage{verbatim}
\usepackage{graphicx}
\usepackage{float}
\usepackage{threeparttable}
\usepackage{bbding}
\usepackage{makecell}
\usepackage{svg}
\usepackage{amssymb}
\usepackage{tabularx} % 引入 tabularx 包
\usepackage{booktabs}
\usepackage{caption}
\usepackage{multirow}
\usepackage{tikz} 
\usepackage{xcolor}
\usepackage{lipsum} % for dummy text
\definecolor{NeedCorrect}{RGB}{255,0,0}

\usetikzlibrary{automata}
\usetikzlibrary{shapes.geometric} 
\usetikzlibrary{positioning,arrows.meta}
\def\BibTeX{{\rm B\kern-.05em{\sc i\kern-.025em b}\kern-.08em
    T\kern-.1667em\lower.7ex\hbox{E}\kern-.125emX}}
    
\begin{document}

\title{Towards Optimal Circuit Generation: Multi-Agent Collaboration Meets Collective Intelligence
\\
{\footnotesize \textsuperscript{*}Note: This work has been submitted to the IEEE for possible publication. Copyright may be transferred without notice.}
}

\author{
\IEEEauthorblockN{
Haiyan Qin\IEEEauthorrefmark{1},
Jiahao Feng\IEEEauthorrefmark{3},  % 修改标记为3
Xiaotong Feng\IEEEauthorrefmark{1},
Wei W. Xing\IEEEauthorrefmark{2}\IEEEauthorrefmark{4},
Wang Kang\IEEEauthorrefmark{1}\IEEEauthorrefmark{4}
}
\IEEEauthorblockA{
\IEEEauthorrefmark{1}
\textit{National Key Laboratory of Spintronics, Hangzhou International Innovation Institute;}\\
\textit{School of Integrated Circuit Science and Engineering, Beihang University, China}
}
\IEEEauthorblockA{  % 更新为武汉理工机电学院
\IEEEauthorrefmark{3}
\textit{School of Mechanical and Electronic Engineering,}\\
\textit{Wuhan University of Technology, Wuhan, China}
}
\IEEEauthorblockA{
\IEEEauthorrefmark{2}
\textit{School of Mathematical and Physical Sciences, University of Sheffield, Sheffield, United Kingdom}
}
\IEEEauthorblockA{
\IEEEauthorrefmark{1}
\{\textit{haiyanq, betty513, wang.kang}\}@buaa.edu.cn,
\IEEEauthorrefmark{3}
325167@whut.edu.cn,
\IEEEauthorrefmark{2}
w.xing@sheffield.ac.uk
}
\thanks{\IEEEauthorrefmark{4}Denotes Corresponding Authors.}
\vspace{-3em}
}

\maketitle

\begin{abstract}

Large language models (LLMs) have transformed code generation, yet their application in hardware design produces gate counts 38\%--1075\% higher than human designs. We present CircuitMind, a multi-agent framework that achieves human-competitive efficiency through three key innovations: syntax locking (constraining generation to basic logic gates), retrieval-augmented generation (enabling knowledge-driven design), and dual-reward optimization (balancing correctness with efficiency). To evaluate our approach, we introduce TC-Bench, the first gate-level benchmark harnessing collective intelligence from the TuringComplete ecosystem---a competitive circuit design platform with hundreds of thousands of players. Experiments show CircuitMind enables 55.6\% of model implementations to match or exceed top-tier human experts in composite efficiency metrics. Notably, our framework elevates the 14B-parameter Phi-4 model (ranked 6th among 1,000 human experts on TuringComplete) to outperform both GPT-4o mini and Gemini 2.0 Flash without requiring specialized training. These innovations establish a new paradigm for hardware optimization where collaborative AI systems leverage collective human expertise to achieve optimal circuit designs. Our model, data, and code are open-source at \url{https://github.com/BUAA-CLab/CircuitMind}.

\end{abstract}


\section{Introduction}

Large language models (LLMs) have demonstrated revolutionary capabilities in code generation across multiple domains, yet their application to hardware design reveals a significant efficiency gap. When tasked with generating digital circuits, even state-of-the-art LLMs produce implementations requiring 38\%--1075\% more gates than those created by skilled human designers. As shown in Table~\ref{tab:comparison_results}, this inefficiency is particularly pronounced in complex circuits involving conditional logic and arithmetic operations.

\begin{table}[h]
    \centering
    \caption{Comparison of LLM-Generated and Human-Optimized Circuits in TuringComplete\cite{TC}}
    \label{tab:comparison_results}
    \vspace*{-2pt}
    \begin{tabular}{@{}lrrr@{}}
    \toprule
    \textbf{Circuit} & \textbf{LLM (Behavioral)} & \textbf{Human Expert} & \textbf{Overhead} \\
    \midrule
    8-bit Adder & 142 & 57-75 & 89\%-149\% \\
    Logic Engine & 125 & 52-70 & 79\%-140\% \\
    Arithmetic Engine & 348 & 102-252 & 38\%-241\% \\
    Conditional Check & 94 & 8-18 & 422\%-1075\% \\
    \bottomrule
    \end{tabular}
    % \vspace*{2pt}
    \footnotesize{Synthesis via Yosys\cite{wolf2016yosys} (LLM: Gemini 2.0 Flash)}
    \vspace*{-2pt}
    \end{table}

This significant performance disparity stems from what we term the \textit{Boolean Optimization Barrier}—a fundamental limitation in how individual LLMs approach gate-level digital design. Unlike human experts who apply systematic minimization techniques and Boolean algebra manipulations, LLMs lack the structured search processes and symbolic reasoning capabilities essential for finding globally optimal solutions at the netlist level. They excel at generating functionally correct code but struggle with the complex trade-offs required to achieve physical efficiency. This barrier becomes particularly evident when analyzing how LLMs handle conditional statements, combinational loops, and arithmetic operations, which they often implement using redundant gate structures that human experts would readily simplify.

% The field's response to this challenge has evolved through three distinct phases (Fig.~\ref{fig:phase_evo}). Initial efforts focused on behavioral code generation with frameworks like VeriGen~\cite{verigen} and RTLCoder~\cite{rtlcoder}, achieving modest functional accuracy but neglecting physical efficiency. The second phase enhanced verification through multi-agent approaches like MAGE~\cite{mage} and AIVRIL~\cite{aivril}, improving correctness but still targeting high-level RTL rather than gate-level optimization. Recent frameworks like BetterV~\cite{betterv} and CodeV~\cite{codev} have begun incorporating physical optimization, yet they continue generating behavioral code that requires synthesis tools to produce netlists, ultimately yielding suboptimal gate-level implementations.

The field's response to this challenge has evolved through three distinct phases (Fig.~\ref{fig:phase_evo}). Initial efforts focused on behavioral code generation, achieving modest functional accuracy while neglecting physical efficiency. The second phase enhanced verification through multi-agent approaches, improving correctness but still targeting high-level RTL rather than gate-level optimization. Recent frameworks have begun incorporating physical optimization, yet continue generating behavioral code that yields suboptimal gate-level implementations.

This evolution mirrors compiler development, progressing from simple translation to sophisticated multi-pass optimization. However, a critical insight remains overlooked: hardware design requires specialized reasoning processes that a single-agent LLM fundamentally cannot replicate. The solution to the Boolean Optimization Barrier lies not in larger models or more domain-specific datasets, but in rethinking the fundamental architecture of AI-based hardware design systems.

% This evolutionary mirrors compiler development, which progressed from simple translation to sophisticated multi-pass optimization. However, a critical insight remains overlooked: just as modern compilers employ multiple specialized passes rather than a single comprehensive algorithm to transform high-level code into efficient executables, hardware design requires specialized reasoning processes that a single-agent LLM fundamentally cannot replicate. The solution to the Boolean Optimization Barrier lies not in creating larger models or more domain-specific datasets, but in rethinking the fundamental architecture of AI-based hardware design systems.
% 
% This insight motivates our paradigm-shifting approach: transitioning from single-agent generative design to multi-agent collaborative optimization. 
% We introduce CircuitMind, a novel framework that distributes complex gate-level reasoning across specialized agents that mirror expert design teams, and TC-Bench, a benchmark that leverages collective intelligence from competitive circuit designers to provide human-aligned evaluation. Together, these innovations establish a new paradigm for hardware optimization where collaborative AI systems overcome individual reasoning limitations through architectural innovation rather than through greater scale or additional training data.

We introduce CircuitMind, a novel framework that distributes complex gate-level reasoning across specialized agents that mirror expert design teams, and TC-Bench, a benchmark that leverages collective intelligence from competitive circuit designers to provide human-aligned evaluation. Together, these innovations establish a new paradigm where collaborative AI systems overcome individual reasoning limitations through architectural innovation rather than through greater scale or additional training data.

Our key contributions include:
\begin{itemize}
\item \textbf{Architectural Innovation:} A hierarchical six-agent framework that overcomes the Boolean Optimization Barrier by distributing complex reasoning tasks across specialized roles, enabling sophisticated gate-level optimization without specialized netlist training.

\item \textbf{Syntax Locking (SL):} A constraint mechanism restricting generation to basic logic gates, eliminating behavioral abstraction and forcing genuine Boolean reasoning at the netlist level.

\item \textbf{Dual-Reward Optimization:} A framework balancing functional correctness with physical efficiency, yielding up to 950\% improvement in Solution Efficiency Index (SEI).

\item \textbf{Human-Competitive Results:} CircuitMind enables 55.6\% of implementations to match or exceed top-tier human experts, with our Phi-4\cite{abdin2024phi} implementation outperforming both GPT-4o mini\cite{achiam2023gpt} and Gemini 2.0 Flash.

\item \textbf{TC-Bench:} The first gate-level benchmark with human-aligned metrics across 28 diverse circuits, establishing realistic efficiency targets based on competitive human design expertise.
\end{itemize}

The integration of CircuitMind with TC-Bench demonstrates that the efficiency gap between AI-generated and human-optimized circuits can be significantly narrowed through intelligent collaboration that overcomes inherent reasoning limitations, rather than through larger models or more training data.

\section{Background and Problem Definition}
Digital circuit design at the gate level requires balancing functional correctness with physical efficiency. We define the problem domain, establish key metrics, analyze AI-based design approaches, and identify the limitations preventing human-competitive efficiency.

\subsection{Gate-Level Circuit Optimization}
A digital circuit can be represented as a directed acyclic graph $G=(V,E)$, where vertices $V$ represent logic gates and edges $E$ represent signal connections. The optimization objective is to minimize both the total gate count $|V|$ and critical path delay $D(G)$, while preserving functionality $F(G)$:
\begin{equation}
\min_{G} \quad \alpha |V| + \beta D(G), \quad \textrm{s.t.} \quad F(G) = F_{\text{spec}},
\end{equation}
where $\alpha,\beta$ are weighting parameters. This NP-hard problem requires non-local reasoning. Traditional logic synthesis tools like Yosys\cite{wolf2016yosys} and ABC~\cite{brayton2010abc} employ heuristic minimization algorithms and technology mapping to tackle this complexity, while human experts apply Boolean algebra manipulation, Karnaugh maps, shared logic identification, and technology mapping across entire circuit structures.   

\subsection{Measuring Design Efficiency}
We define the Solution Efficiency Index (SEI) for functionally correct implementations:
\begin{equation}
\label{eq:sei_task}
\text{SEI}_{\text{task}} = \frac{1}{\alpha \cdot G + \beta \cdot D}
\end{equation}
where $G$ represents gate count and $D$ represents delay. We use $\alpha = \beta = 1$ for balanced evaluation, reflecting common optimization goals in TuringComplete~\cite{TC}. However, real-world circuit optimization often involves complex trade-offs between area, delay, and power, and the TuringComplete~\cite{TC} scoring mechanism might incorporate additional factors like simulation ticks. For benchmark-level aggregation, we use geometric mean:
\begin{equation}
\label{eq:sei_benchmark}
\text{SEI}_{\text{benchmark}} = \exp\left( \frac{1}{n} \sum_{i=1}^n \ln(\max(\text{SEI}_{\text{task}, i}, \epsilon)) \right)
\end{equation}
% where $n$ is the total number of tasks, $\text{SEI}_{\text{task},i}$ is the Task-Level SEI for task $i$ (0 if failed), and $\epsilon$ is a small positive constant ensuring the logarithm is defined for failed tasks.
where $n$ is the total number of tasks, $\text{SEI}_{\text{task},i}$ is the SEI for task $i$ (0 if failed), and $\epsilon = 10^{-5}$ ensures that the logarithm is defined for failed tasks.

\subsection{Evolution of AI in Hardware Design}
\begin{figure}[t]
\centering
\includegraphics[width=0.45\textwidth]{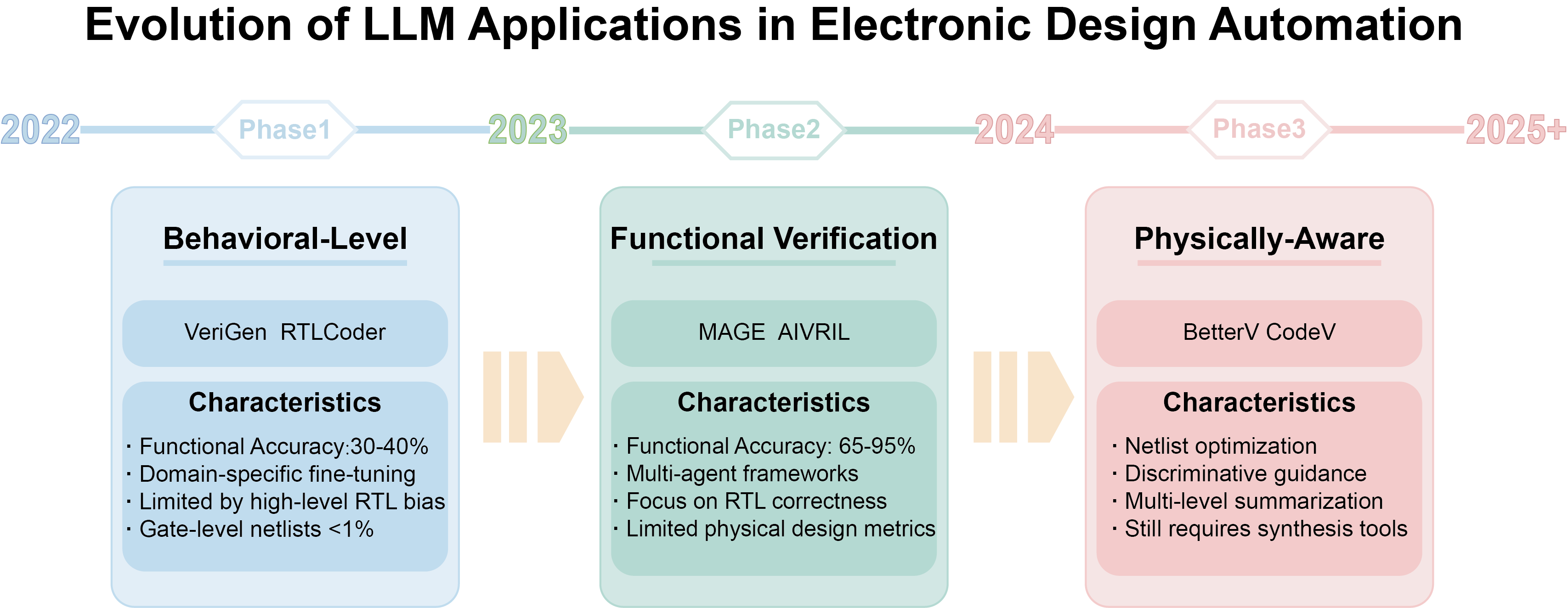}
\caption{LLM-based EDA Evolution: From Behavioral to Physical Design}
\label{fig:phase_evo}
\vspace{-1em}
\end{figure}

LLMs in electronic design automation have evolved through three distinct phases (Fig.~\ref{fig:phase_evo}):   

\subsubsection{Phase 1: Behavioral Code Generation (2022-2023)}
Early frameworks like VeriGen~\cite{verigen} and RTLCoder~\cite{rtlcoder} achieved 30-40\% functional accuracy on VerilogEval~\cite{verilogeval} but disregarded physical efficiency. Analysis of PyraNet~\cite{PyraNet} shows gate-level netlists constitute less than 1\% of available HDL code, limiting models' exposure to optimized patterns.   

\subsubsection{Phase 2: Verification-Enhanced Generation (2023-2024)}
Multi-agent frameworks such as MAGE~\cite{mage} and AIVRIL~\cite{aivril} increased accuracy to 65-95\% through verification feedback but operated exclusively at the behavioral RTL level without addressing physical optimization challenges.

\subsubsection{Phase 3: Early Physical Optimization (2024-present)}
Recent approaches like BetterV~\cite{betterv} and CodeV~\cite{codev} incorporate physical considerations but remain limited by generating behavioral code that synthesis tools must convert to gates. As Table~\ref{tab:framework_comparison} shows, while frameworks like RTLSquad~\cite{wang2025rtlsquad} use multi-agent systems for RTL optimization and GenEDA~\cite{fang2025geneda} focuses on reasoning about netlists, no existing framework enforces direct gate-level reasoning or aligns with human expert optimization practices in the same manner as CircuitMind. The qualitative assessments in Table 2 reflect the primary focus of each framework; for instance, BetterV\cite{betterv} uses a discriminator for optimization guidance, while HiVeGen~\cite{hivegen} targets PPA but still operates at the RTL level.   

\begin{table}[t]
    \centering
    \caption{Comparative Analysis of Verilog Generation Frameworks}
    \label{tab:framework_comparison}
    \resizebox{\columnwidth}{!}{%
    \begin{tabular}{lllll}
        \toprule
        \textbf{Framework} & \textbf{Methodology} & \textbf{Gate-Level} & \textbf{PPA} & \textbf{Human} \\
        & & \textbf{Enforcement} & \textbf{Optimization} & \textbf{Alignment} \\
        \midrule
        MAGE \cite{mage} & Multi-Agent & No & Low & No \\
        BetterV \cite{betterv} & Discriminator  & No & Medium & No \\
        CodeV \cite{codev} & Multi-level & No & Low & No \\
        AIVRIL \cite{aivril} & Verification & No & Low & No \\
        PromptV \cite{promptv} & Teacher-Learner & No & Low & No \\
        HiVeGen \cite{hivegen} & Hierarchical & No & High & No \\
        RTLSquad \cite{wang2025rtlsquad}  & Multi-Agent (RTL Opt.) & No & High & No \\
        GenEDA \cite{fang2025geneda}  & Encoder-Decoder (Netlist Reasoning) & No & Low & No \\
        \midrule
        \textbf{CircuitMind} & \textbf{Multi-Agent} & \textbf{Yes (SL)} & \textbf{High (DR)} & \textbf{Yes (TC-Bench)} \\
        \bottomrule
    \end{tabular}%
    }
    \vspace{-1em}
\end{table}   

\subsection{The Boolean Optimization Barrier}
LLM-generated circuits exhibit gate counts 38\%--1075\% higher than human-optimized designs due to four fundamental limitations:

\subsubsection{Associative vs. Structured Reasoning}
LLMs generate sequences based on statistical patterns rather than performing the structured symbolic reasoning required for Boolean optimization. Human experts apply systematic minimization techniques like Quine-McCluskey algorithms\cite{quine1952problem} that LLMs cannot effectively replicate.

\subsubsection{Local vs. Global Optimization}
The autoregressive nature of LLMs forces local, token-by-token decision making without the ability to revise earlier decisions when later improvements become apparent. Circuit optimization requires global restructuring and non-local transformations that are challenging for current generative models.

\subsubsection{Abstraction Leakage}
LLMs naturally gravitate toward behavioral abstractions like \texttt{if-else} statements that introduce significant inefficiencies when synthesized to gate-level netlists, bypassing essential Boolean optimization.

\subsubsection{Limited Training on Optimized Netlists}
The extreme scarcity of gate-level examples in training data severely limits LLMs' exposure to optimized circuit patterns and structures.   

\subsection{Benchmark Misalignment}
% A key challenge in advancing LLM-based hardware design is the disconnect between evaluation metrics and physical implementation quality. As shown in Table~\ref{tab:benchmark_comparison}, early benchmarks like VerilogEval~\cite{verilogeval} primarily focused on functional correctness, neglecting crucial Power, Performance, and Area (PPA) metrics. While subsequent benchmarks like RTLLM~\cite{rtllm} began incorporating limited PPA analysis at the RTL level, they often lacked comprehensive human expert references. More recently, benchmarks like ResBench\cite{resbench} and GenBen have emerged to specifically address these gaps by evaluating PPA/QoR metrics, using diverse data sources, and employing techniques to prevent data contamination. ResBench\cite{resbench} focuses particularly on FPGA resource efficiency, while GenBen aims for broader coverage including synthesizability and debugging.

A key challenge in advancing LLM-based hardware design is the disconnect between evaluation metrics and physical implementation quality. As shown in Table~\ref{tab:benchmark_comparison}, existing benchmarks have evolved from functional correctness (VerilogEval~\cite{verilogeval}) to limited PPA metrics (RTLLM~\cite{rtllm}) and more comprehensive quality assessments (ResBench\cite{resbench}, GenBen\cite{genben}), but most lack alignment with human expert performance. TC-Bench addresses this gap by leveraging collective intelligence from TuringComplete's\cite{TC} competitive environment to establish meaningful gate-level optimization targets based on actual human performance tiers.

Despite these advancements, TC-Bench offers a unique contribution by harnessing collective intelligence from a competitive gaming environment to establish gate-level optimization challenges and human performance tiers based on the SEI metric. This approach provides a distinct perspective focused on direct gate-level optimization efficiency derived from a large community actively competing on such tasks. Table~\ref{tab:benchmark_comparison} provides an updated comparison.

\begin{table*}[t]
    \centering
    \caption{Comparison of Benchmarks for LLMs in Hardware Design}
    \label{tab:benchmark_comparison}
    \renewcommand{\arraystretch}{1.2}
    \setlength{\tabcolsep}{10pt}
    \begin{tabularx}{\textwidth}{@{}lcccl>{\raggedright\arraybackslash}X@{}}
        \toprule
        \textbf{Benchmark} & \textbf{Design Level} & \textbf{Key Metrics} & \textbf{Data Source} & \textbf{Alignment} & \textbf{Key Features} \\
        \midrule
        TC-Bench(ours) & Gate & Func. Corr., SEI & Game (TuringComplete) & Yes (Tiers) & Intelligence source, SEI metric \\
        VerilogEval\cite{verilogeval}  & RTL & Func. Corr. & Academic/Web & Limited & Focus on functional correctness \\
        RTLLM\cite{rtllm}  & RTL & Func. Corr., PPA & Academic/Web & Limited & PPA metrics, reference designs \\
        ResBench\cite{resbench} & RTL & Func. Corr., FPGA & Synthetic/Academic & No & FPGA resource efficiency \\
        GenBen\cite{genben} & Mixed & Func. Corr., QoR, Debug & Diverse (Silicon+) & Limited & QoR focus, contamination handling \\
        RTL-Repo\cite{rtl-repo}  & Multi-file RTL & Func. Corr. & GitHub & Limited & Focus on multi-file projects \\
        \bottomrule
    \end{tabularx}
    \vspace{4pt}
    \begin{tablenotes}[para,flushleft]
    \small
    \textbf{Abbr.:} Func. Corr. = Functional Correctness, 
 FPGA = Field-Programmable Gate Array, QoR = Quality of Results, 

    \end{tablenotes}
    \vspace{-1em}
\end{table*}

\subsection{The Fundamental Reasoning Gap}
Existing approaches attempt to bridge the efficiency gap through three primary strategies, all facing significant limitations. Scale-based approaches fail because the barrier is fundamentally algorithmic rather than capacity-related; simply increasing model size does not address the underlying reasoning limitations in Boolean optimization. Fine-tuning strategies are hampered by extreme scarcity of optimized gate-level examples, creating an insurmountable data bottleneck. Single-agent verification loops, while improving functional correctness, lack the specialized reasoning capabilities necessary for effective Boolean optimization and cannot replicate the structured symbolic manipulation that human experts employ. 

Addressing this gap requires a paradigm shift from scaling individual models to collaborative systems that distribute specialized reasoning tasks across multiple agents—an approach that addresses the fundamental limitations in how LLMs process logical optimization problems.

\section{TC-Bench: Harnessing Collective Intelligence}
\label{sec:tcbench}

To address the benchmark misalignment problem identified earlier, we introduce TC-Bench—a gate-level evaluation framework that leverages collective intelligence from competitive circuit designers on the TuringComplete platform. This framework establishes meaningful efficiency targets for AI-generated circuits based on human performance benchmarks.

\subsection{Collective Intelligence Approach}

\begin{figure}[t]
\centering
\includegraphics[width=0.45\textwidth]{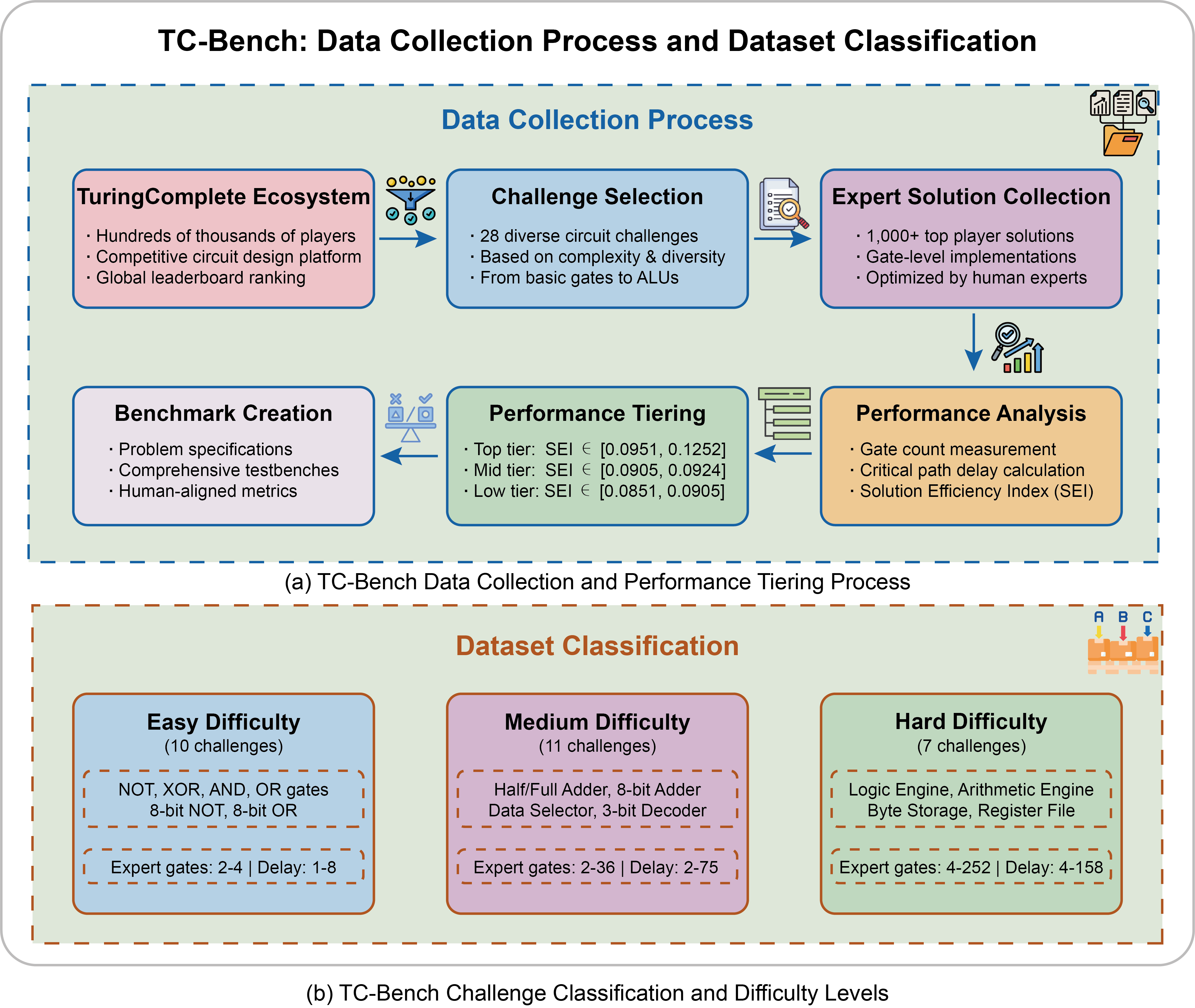}
\caption{TC-Bench's collective intelligence harvesting process}
\label{fig:collective}
\vspace{-1em}
\end{figure}

TC-Bench derives its data from TuringComplete, where a large community competes to optimize digital circuits using minimal logic gates while meeting functional requirements. Solutions are ranked based on gate count and performance (delay), generating thousands of expert-refined implementations that often significantly exceed the efficiency of conventional synthesis tool outputs. TC-Bench harvests these optimized designs across player skill levels (Fig.~\ref{fig:collective}(a)), creating a multi-tiered reference framework that provides diverse, human-optimized implementations rather than relying on standard synthesis results.

% : (a) Data extraction from TuringComplete leaderboards across skill tiers. (b) Benchmark stratification into Easy, Medium, and Hard challenges.
\subsection{Benchmark Structure}

TC-Bench comprises 28 challenges selected based on circuit complexity, functional diversity, and optimization potential, stratified into three difficulty levels (Fig.~\ref{fig:collective}(b)):

\begin{itemize}
    \item \textbf{Easy} (10 challenges): Basic combinational logic implementing fundamental gates and simple selectors, typically requiring 2-4 gates in expert solutions.
        
    \item \textbf{Medium} (11 challenges): More complex components including adders, multiplexers, and basic ALUs, requiring 5-36 gates in expert solutions.
        
    \item \textbf{Hard} (7 challenges): Advanced designs encompassing multi-bit arithmetic, logic engines, and sequence detectors, requiring 40-250+ gates in expert solutions.
\end{itemize}

Each challenge includes a functional specification, benchmarking data from top human designs, and a comprehensive testbench. Crucially, the benchmark requires direct netlist construction using only predefined basic logic gates, enforcing genuine Boolean optimization rather than relying on higher-level HDL abstractions.

\subsection{Human-Aligned Evaluation Metrics}

Building on the evaluation metrics introduced in Section 2.2, TC-Bench employs the Solution Efficiency Index (SEI) to compare AI-generated circuits against human expertise.

\subsubsection{Human Reference Tiers}

To enable nuanced comparison against human expertise, we analyzed the aggregated SEI scores from TuringComplete leaderboard ranks, identifying three representative performance tiers:

\begin{itemize}
    \item \textbf{Top tier} (Rank 1-10): Expert optimizers. SEI range: [0.0951, 0.1252].
    \item \textbf{Mid tier} (Rank 11-300): Skilled practitioners. SEI range: [0.0905, 0.0924].
    \item \textbf{Low tier} (Rank 301-1000): Competent designers. SEI range: [0.0851, 0.0905].
\end{itemize}

These tiers establish concrete targets for LLM-based circuit generation frameworks and provide a meaningful scale for measuring progress toward human-competitive efficiency. By positioning AI systems on this human-aligned scale, TC-Bench enables direct comparison between machine-generated and expert-optimized implementations in terms that matter to real-world hardware design.

\section{CircuitMind: Multi-Agent Collaboration}

To overcome the Boolean Optimization Barrier in circuit design, we introduce CircuitMind—a hierarchical multi-agent framework that distributes complex reasoning tasks across specialized agents, enabling capabilities that individual LLMs cannot achieve alone.

\subsection{Architectural Overview}

CircuitMind adopts a three-tier architecture inspired by engineering team hierarchies, with six specialized agents distributed across strategic, coordination, and execution layers:

\begin{figure}[t]
\centering
\includegraphics[width=0.45\textwidth]{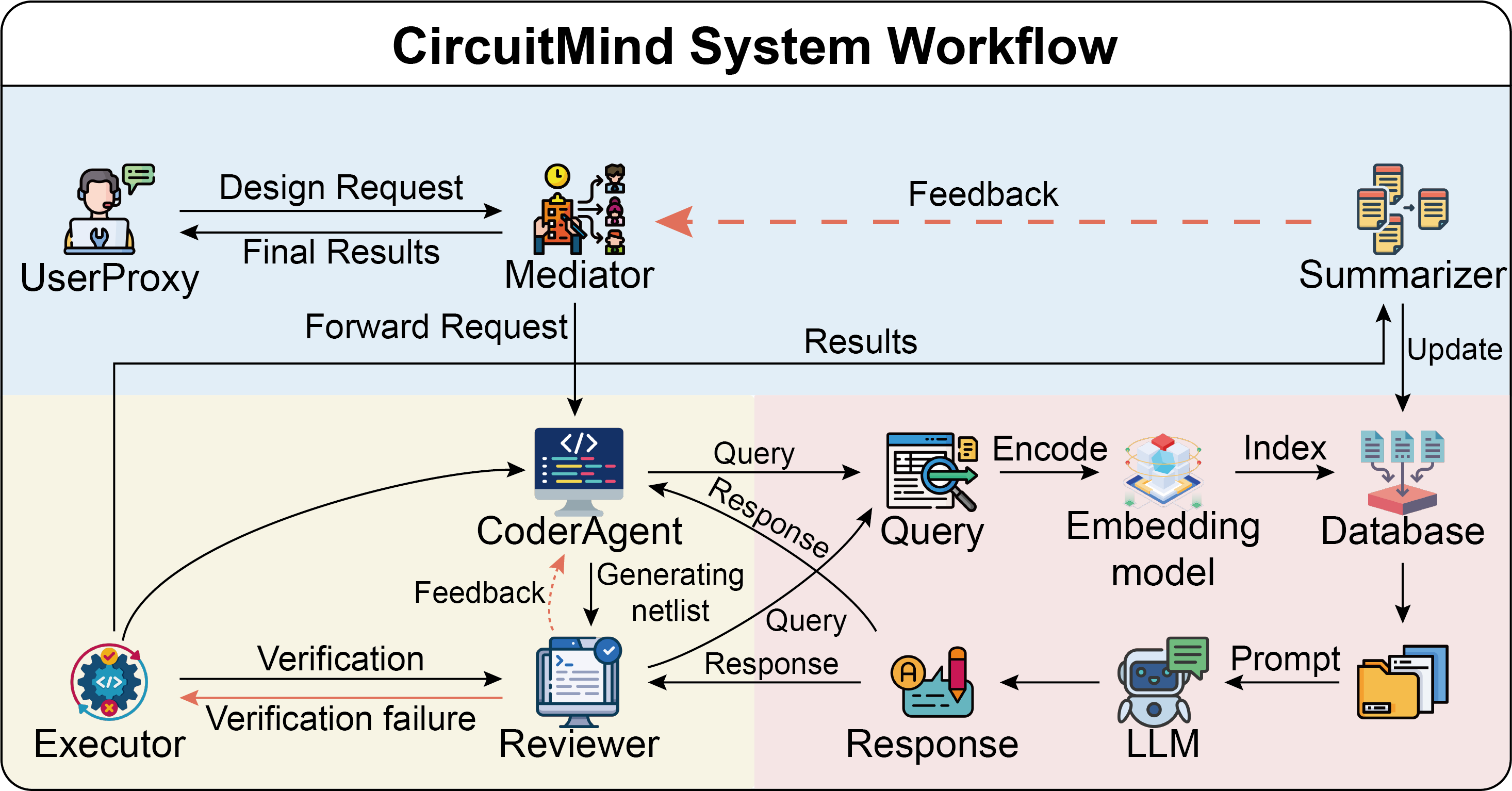}
\caption{CircuitMind System Architecture}
\label{fig:architecture}
\vspace{-1em}
\end{figure}

\begin{itemize}
    \item \textbf{Strategic Layer:} UserProxy (translates requirements into formal specifications) and Mediator (orchestrates agent interactions and resource allocation).
    
    \item \textbf{Coordination Layer:} Reviewer (provides PPA-focused feedback) and Summarizer (extracts and archives optimization patterns).
    
    \item \textbf{Execution Layer:} CoderAgent (generates gate-level netlists) and Executor (performs verification through compilation and simulation).
\end{itemize}

This structured decomposition addresses the Boolean optimization challenge by distributing specialized reasoning tasks across multiple agents, enabling capabilities that a single LLM cannot achieve. Each agent focuses on a specific aspect of the design process—specification analysis, code generation, verification, or optimization—allowing the system to overcome individual limitations through collaborative problem-solving.

\subsection{Core Technical Innovations}

CircuitMind introduces three key innovations that directly address the fundamental limitations of LLMs in circuit design:

\subsubsection{Syntax Locking}

Syntax Locking (SL) constrains generation to five basic logic gates (AND, OR, NOT, XOR, NAND), enforcing genuine Boolean optimization and preventing behavioral abstraction leakage. By implementing a strict context-free grammar limited to fundamental combinational elements, SL forces models to perform actual Boolean expression simplification rather than relying on high-level abstractions.

This constraint parallels the environment faced by human designers in TuringComplete, where optimization occurs through strategic arrangement of basic components. SL is enforced through multiple mechanisms: explicit netlist format requirements in prompts, retrieval of optimized netlist-level examples, and Reviewer verification of gate-level compliance.

\subsubsection{Retrieval-Augmented Generation}

Retrieval-Augmented Generation (RAG)\cite{RAG} enables knowledge-driven circuit design by identifying and reusing optimized subcircuits, allowing efficient learning from limited examples. The implementation uses a dynamic knowledge database where successful circuit patterns are stored, indexed by functionality, and retrieved during the design process.

When receiving a design task, CoderAgent first classifies the circuit type (combinational or sequential) and identifies required building blocks. It then queries the knowledge database for optimized implementations of these components, incorporating them directly into generation prompts as concrete examples of efficient design patterns. This approach enables the model to leverage past successes while adapting to new requirements, creating a bootstrapping mechanism that progressively enhances design capabilities.

The knowledge database evolves as CircuitMind successfully completes design tasks, with Summarizer extracting optimized subcircuits and adding them to the knowledge base. Initially populated with only basic logic gates, the database gradually accumulates more sophisticated modules, enabling the system to tackle increasingly complex designs without requiring extensive pre-training on gate-level netlists.

\subsubsection{Dual-Reward Optimization}

Dual-Reward Optimization explicitly balances functional correctness with physical efficiency metrics through dynamic feedback loops. This framework combines two complementary evaluation dimensions:

Functional Correctness Score measures how well the generated netlist meets specified requirements through comprehensive testing, quantified as the percentage of test cases passed.

Physical Efficiency Score evaluates gate count, critical path delay, and other physical design metrics against reference implementations, calculated as a weighted sum adjusted according to design priorities.

These dual rewards guide the generation process through Reviewer feedback targeting both functional issues and inefficient structures, CoderAgent prioritization of components with demonstrated good PPA metrics, and knowledge database rankings based on efficiency metrics. This approach creates explicit optimization incentives that mimic the competitive pressures in the TuringComplete ecosystem, guiding models toward designs that not only work correctly but approach human expert efficiency.

\begin{figure}[t]
\centering
\includegraphics[width=0.45\textwidth]{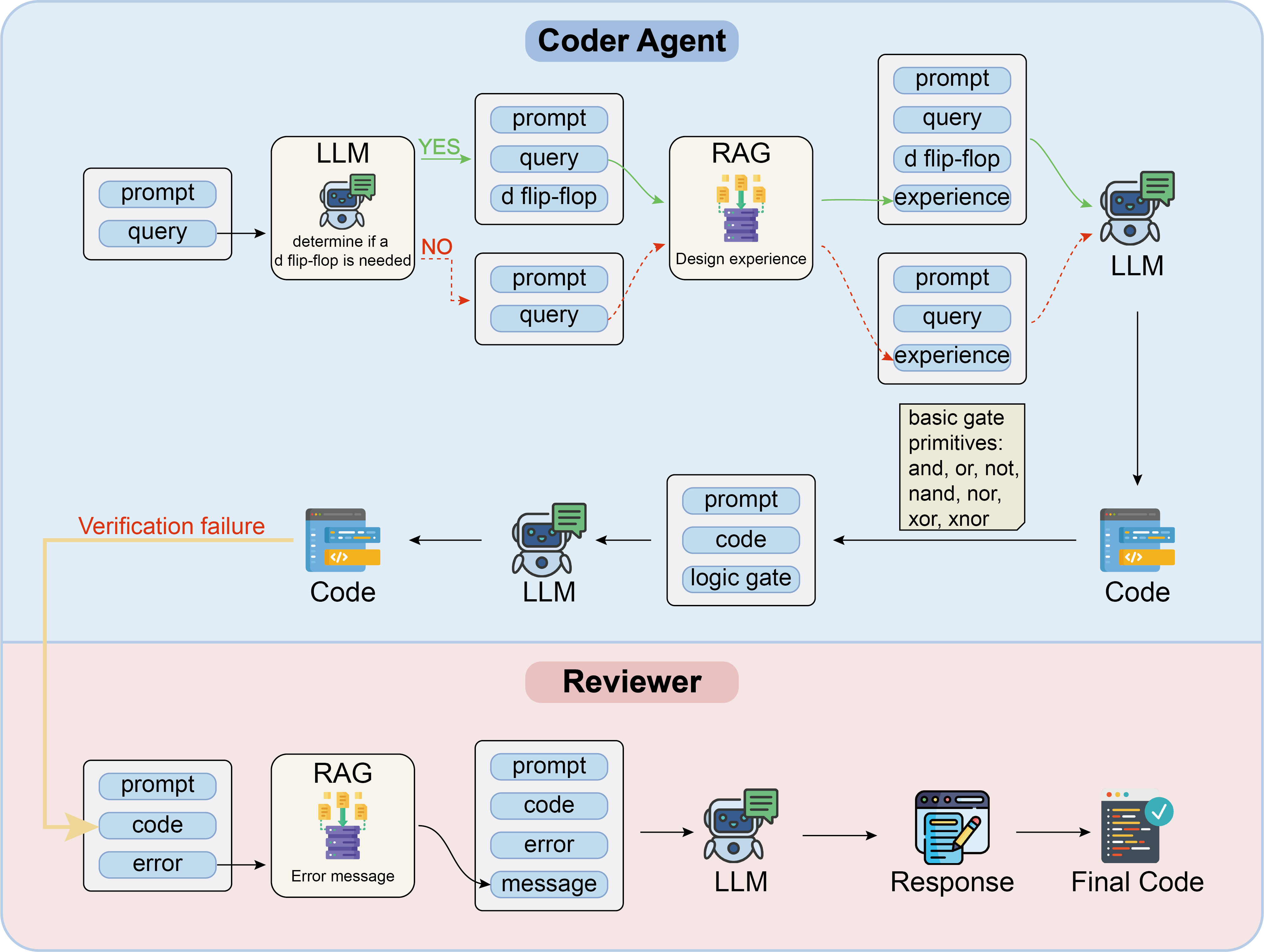}
\caption{CoderAgent and Reviewer dynamic prompting process}
\label{fig:prompt_generation}
\vspace{-1em}
\end{figure}

\subsection{Collaborative Workflow}

CircuitMind's collaborative workflow (Fig.~\ref{fig:architecture}) begins when UserProxy receives a circuit design task and converts it into a structured specification. Mediator routes this specification to CoderAgent, which initiates a detailed analysis process:

First, CoderAgent determines whether the circuit requires combinational or sequential logic, guiding which netlist templates will be retrieved from the knowledge database.

Next, it analyzes overall functionality to identify required building blocks, querying the knowledge database for netlist-level representations of these components.

CoderAgent then generates a complete circuit implementation incorporating these optimized components while adhering to syntax locking constraints.

Mediator routes the generated netlist to Reviewer, which performs two-step verification: confirming adherence to netlist-level requirements and evaluating physical efficiency metrics.

Executor performs functional verification through compilation and simulation, with results fed back to Reviewer for potential optimization. This dynamic prompting process between CoderAgent and Reviewer is illustrated in Fig.~\ref{fig:prompt_generation}.

If verification succeeds, Summarizer updates the knowledge database with newly validated optimized patterns; if errors occur, Reviewer queries the knowledge database for similar error patterns and potential solutions, initiating an iterative debugging process.

This structured collaboration addresses the Boolean optimization challenge by decomposing complex reasoning tasks across specialized agents. While individual LLMs struggle with maintaining complex state across optimization steps and evaluating PPA tradeoffs, CircuitMind's distributed architecture enables sophisticated reasoning through agent specialization and coordinated interaction.

\subsection{Knowledge Evolution Mechanism}

CircuitMind achieves continuous improvement through a dual-layer learning architecture:

Short-Cycle Design Refinement involves Reviewer suggesting modifications based on knowledge database insights, CoderAgent adopting optimized solutions upon successful verification, and Summarizer logging effective strategies. This immediate feedback loop enables rapid adaptation within individual design tasks, with agents learning from both successes and failures in real-time.

Long-Term Knowledge Evolution involves systematically capturing optimization patterns in a structured knowledge base, allowing identification of common design patterns and proactive optimization in subsequent projects. The knowledge acquisition process encompasses experience capture (extracting optimization patterns from successful designs), knowledge organization (indexing information by functional requirements and error types), and adaptive reuse (applying historical solutions to new challenges).

This evolutionary approach allows CircuitMind to overcome data scarcity limitations by creating its own internal optimization targets based on accumulated design experience. Rather than requiring extensive gate-level training data, the system bootstraps its capabilities through iterative refinement and knowledge accumulation, becoming increasingly capable with each successfully completed design task.

\section{Experimental Results}

To evaluate CircuitMind's effectiveness in generating gate-level digital circuits, we conducted comprehensive experiments using the TC-Bench benchmark, comparing various model configurations against human expert performance tiers.

\subsection{Experimental Setup}

We evaluated CircuitMind using a diverse set of LLMs as backend engines: open-source models (Qwen series\cite{yang2024qwen2}, Phi-4\cite{abdin2024phi}, DeepSeek series\cite{liu2024deepseek}\cite{guo2025deepseek}) and commercial models (GPT-4o mini\cite{achiam2023gpt}, Gemini 2.0 Flash\cite{Gemini}). These models powered the core CoderAgent and Reviewer components, while Qwen2.5:14B was consistently used for RAG components. For baseline comparison, we used direct circuit generation by the same LLMs without the CircuitMind framework. To ensure fair comparison, we applied consistent conditions across all experiments: 4-bit quantization for most open-source models (except DeepSeek), API access for commercial and DeepSeek models, and a maximum of 2 revision attempts per task.

We evaluated CircuitMind using the Pass@k metric from \cite{chen2021evaluating}, defined as:

\begin{equation}
\text{Pass@k} = \mathbb{E}_{\text{Problems}}\left[1 - \frac{\binom{n-c}{k}}{\binom{n}{k}}\right],
\end{equation}

where $k$ is samples per task, $n$ is total samples, and $c$ is correct solutions. With $n=20$, we estimated correctness from correct sample fractions. Pass@k reflects the probability of solving a task with the best of $k$ samples.

All implementations were synthesized to gate-level netlists using Yosys 0.34~\cite{wolf2016yosys} with standard optimization passes disabled, ensuring results reflect the actual logical structures produced by the models rather than the synthesis tool's capabilities.

% CircuitMind was implemented using diverse open-source (Qwen series\cite{yang2024qwen2}, Phi-4\cite{abdin2024phi}, DeepSeek series\cite{liu2024deepseek}\cite{guo2025deepseek}) and commercial (GPT-4o mini\cite{achiam2023gpt}, Gemini 2.0 Flash\cite{Gemini}) LLMs as backends for core generation and review agents (CoderAgent, Reviewer), with Qwen2.5:14B consistently used for RAG components (DesignRAG, ReviewRAG). Baselines involved direct generation by the LLMs. Open-source models (except DeepSeek) were 4-bit quantized. Commercial and DeepSeek models used APIs. A limit of 2 revision attempts per task ensured fair comparison.

% All generated Verilog code was synthesized into gate-level netlists using Yosys 0.34~\cite{wolf2016yosys}. Crucially, standard synthesis optimizations (`opt` pass in Yosys) were disabled to directly reflect the logical structure generated by the LLM or framework, rather than the capabilities of the synthesis tool itself. 
% % Removed potentially incorrect Icarus citation

\subsection{Comparative Performance Analysis}

Table~\ref{tab:model_results} presents the comprehensive performance comparison between the baseline LLMs and their CircuitMind-enhanced counterparts (CM) across the easy, medium, and hard tasks defined in TC-Bench, evaluated using Pass@1 and the SEI.

\begin{table}[t]
    \centering
    \caption{TC-Bench Performance: Base LLMs vs. CircuitMind}
    \label{tab:model_results}
    \resizebox{\columnwidth}{!}{%
    \begin{tabular}{lcccccccc}
        \toprule
        \multirow{2}{*}{Model} & \multicolumn{4}{c}{Pass@1} & \multicolumn{4}{c}{SEI} \\
        \cmidrule(lr){2-5} \cmidrule(lr){6-9}
         & Easy & Medium & Hard & Overall & Easy & Medium & Hard & Overall \\
        \midrule
        Gemini 2.0 Flash & 0.60 & 0.53 & 0.11 & 0.434 & 0.21 & 0.06 & 0.02 & 0.063 \\
        CM(Gemini-2.0) & 0.97 & 0.76 & 0.94 & 0.888 & 0.17 & 0.09 & 0.07 & \textbf{0.102} \\
        \midrule
        GPT-4o mini & 0.86 & 0.64 & 0.13 & 0.570 & \textbf{0.30} & 0.07 & 0.001 & 0.028 \\
        CM(GPT-4o mini) & 0.89 & 0.64 & 0.82 & 0.779 & 0.14 & 0.09 & 0.09 & \textbf{0.104} \\
        \midrule
        DS-R1 & \textbf{1.00} & \textbf{0.96} & 0.26 & 0.775 & 0.16 & 0.06 & 0.01 & 0.046 \\
        CM(DS-R1) & \textbf{1.00} & 0.84 & 0.99 & \textbf{0.936} & 0.13 & 0.09 & 0.03 & \textbf{0.071} \\
        \midrule
        DS-V3 & 0.98 & 0.82 & 0.11 & 0.671 & 0.23 & 0.09 & 0.002 & 0.035 \\
        CM(DS-V3) & \textbf{1.00} & 0.82 & 0.99 & \textbf{0.929} & 0.11 & 0.04 & 0.03 & \textbf{0.051} \\
        \midrule
        Phi-4 & 0.67 & 0.38 & 0.02 & 0.380 & 0.25 & 0.07 & 0.001 & 0.026 \\
        CM(Phi-4) & 0.95 & 0.61 & 0.26 & 0.629 & 0.17 & 0.10 & \textbf{0.089} & \textbf{0.115} \\
        \midrule
        QW:7B & 0.45 & 0.29 & 0.00 & 0.263 & 0.21 & 0.04 & 0.00 & 0.010 \\
        CM(QW:7B) & 0.49 & 0.22 & 0.13 & 0.288 & 0.15 & \textbf{0.13} & 0.06 & \textbf{0.105} \\
        \midrule
        QW:14B & 0.64 & 0.36 & 0.00 & 0.355 & 0.28 & 0.04 & 0.00 & 0.010 \\
        CM(QW:14B) & 0.93 & 0.36 & 0.14 & 0.500 & 0.14 & 0.06 & 0.08 & \textbf{0.088}\\
        \midrule
        QWC:14B & 0.53 & 0.38 & 0.06 & 0.341 & 0.18 & 0.06 & 0.001 & 0.022 \\
        CM(QWC:14B) & 0.80 & 0.38 & 0.28 & 0.500 & 0.19 & 0.08 & 0.06 & \textbf{0.097} \\
        \midrule
        QWC:32B & 0.67 & 0.36 & 0.04 & 0.377 & 0.27 & 0.07 & 0.001 & 0.027 \\
        CM(QWC:32B) & 0.99 & 0.62 & 0.32 & 0.664 & 0.16 & 0.09 & 0.05 & \textbf{0.090} \\
        \bottomrule
    \end{tabular}%
    }

    \footnotesize{CM: CircuitMind; DS: DeepSeek; QW: Qwen2.5; QWC: Qwen2.5-Coder. SEI values are averaged across tasks in each category.}
    \vspace{-1em}
\end{table}

\begin{figure}[t]
\centering
% Assuming the image file exists at this path
\includegraphics[width=0.95\columnwidth]{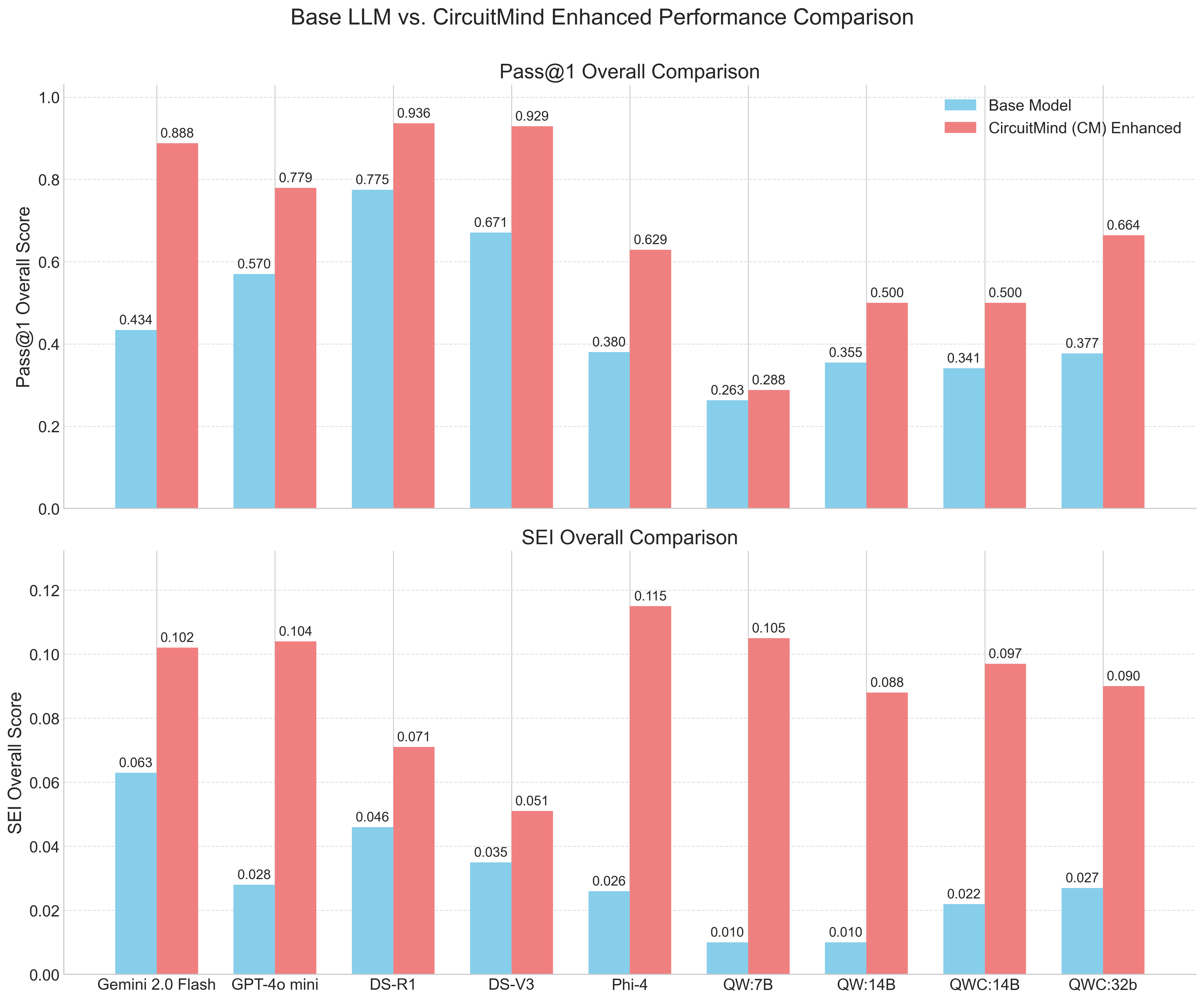}
\caption{Overall Pass@1 and SEI Comparison: Base LLMs vs. CircuitMind.}
\label{fig:Pass1_SEI_comparison}
\vspace{-1em}
\end{figure}

 The results, detailed in Table~\ref{tab:model_results} and summarized in Fig.~\ref{fig:Pass1_SEI_comparison}, reveal key patterns analyzed below.

% --- Placement for Figure 1 (Pass@1 and SEI Overall Comparison) ---
% --------------------------------------------------------------------

\subsubsection{Functional Correctness and Physical Efficiency Enhancements}
CircuitMind delivers substantial improvements in both functional correctness and physical efficiency across all evaluated models. For functional correctness (Fig.~\ref{fig:Pass1_SEI_comparison}, top), the most striking gains appear on hard tasks, where baseline LLMs typically struggled. Notable improvements include DeepSeek-R1 (0.26 to 0.99, $3.8\times$ increase) and Phi-4 (0.02 to 0.26, $13\times$ increase) on hard tasks.

For physical efficiency (Fig.~\ref{fig:Pass1_SEI_comparison}, bottom), CircuitMind consistently improves overall SEI for \textbf{all} tested models. The efficiency gains are particularly significant on hard tasks, with some models showing order-of-magnitude improvements: CM(GPT-4o mini) achieved $90\times$ higher SEI and CM(Phi-4) $89\times$ higher SEI compared to their respective baselines. While some models showed slight SEI decreases on easy tasks, the substantial improvements on medium and hard tasks resulted in significant overall efficiency gains, demonstrating CircuitMind's effectiveness in complex optimization scenarios.

\subsubsection{Model-Specific Performance Patterns}
Performance gains varied notably across different models, revealing interesting patterns in framework-model compatibility. CM(Phi-4) achieved the highest overall SEI (0.115), representing a 342\% increase over its baseline—the largest relative improvement among all tested models. Commercial models also showed substantial gains, with CM(GPT-4o mini) and CM(Gemini-2.0) increasing overall SEI by 271\% and 61.9% respectively.

Interestingly, performance improvements were not uniform across difficulty levels. For instance, CM(GPT-4o mini) showed a 53.3\% SEI decrease on easy tasks but dramatic improvements on harder tasks, resulting in an overall 271\% SEI gain. This suggests that CircuitMind's effectiveness varies based on how well each model's inherent strengths align with the framework's collaborative optimization approach.

\begin{figure}[t]
\centering
% Assuming the image file exists at this path
\includegraphics[width=0.95\columnwidth]{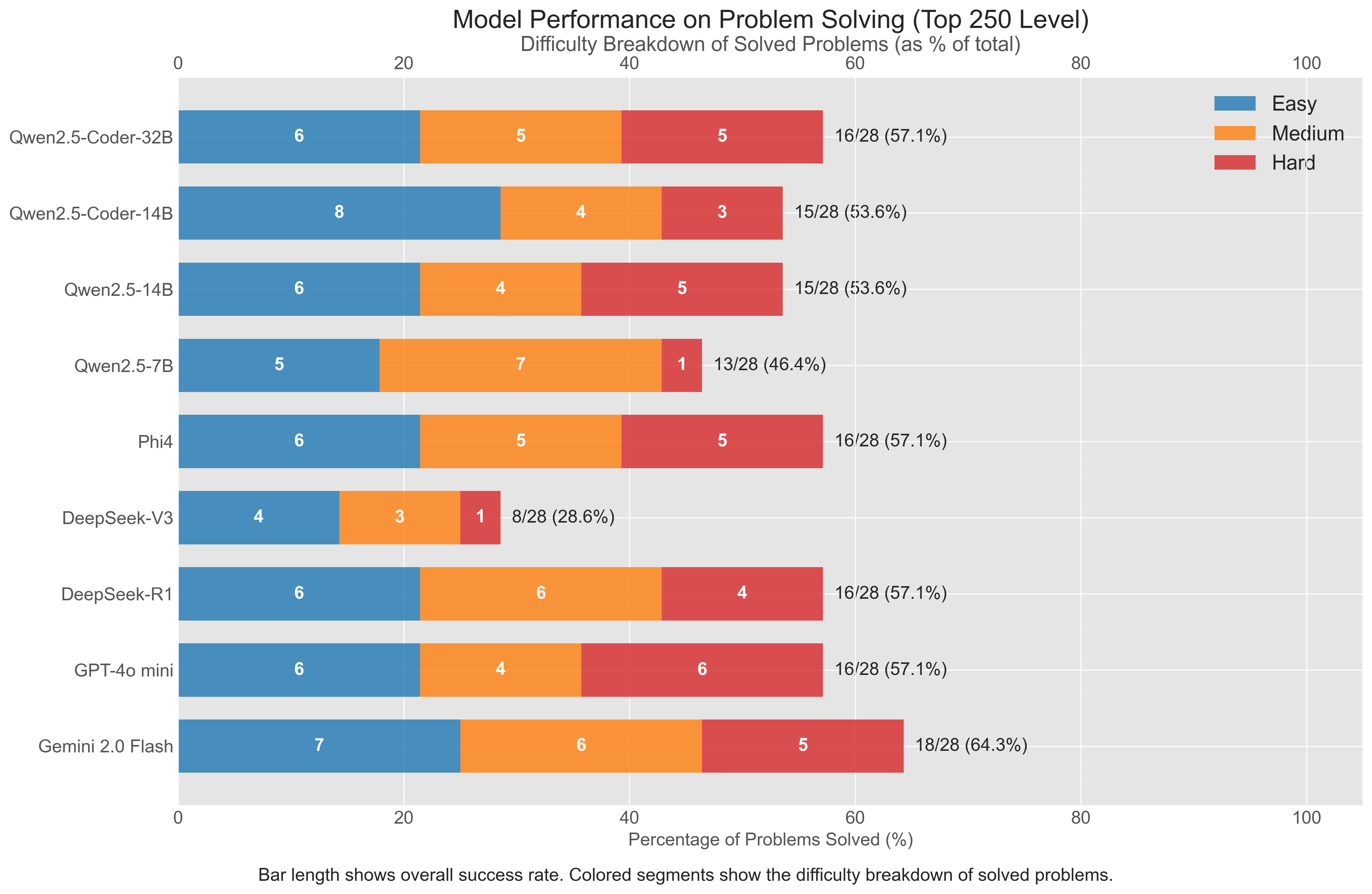}
\caption{Solved Problem Count by Difficulty.}
\label{fig:difficulty_breakdown}
\vspace{-1em}
\end{figure}

Fig.~\ref{fig:difficulty_breakdown} confirms these observations, showing that high-performing implementations like CM(Gemini-2.0) and CM(DeepSeek-R1/V3) successfully solve problems across all difficulty categories, with particularly strong performance on hard tasks that baseline models struggled to address.

% --- Placement for Figure 2 (Difficulty Breakdown) ---
% -----------------------------------------------------

\subsection{Human-Competitive Performance Evaluation}
\begin{figure}[t]
\centering
% Assuming the image file exists at this path
\includegraphics[width=0.9\columnwidth]{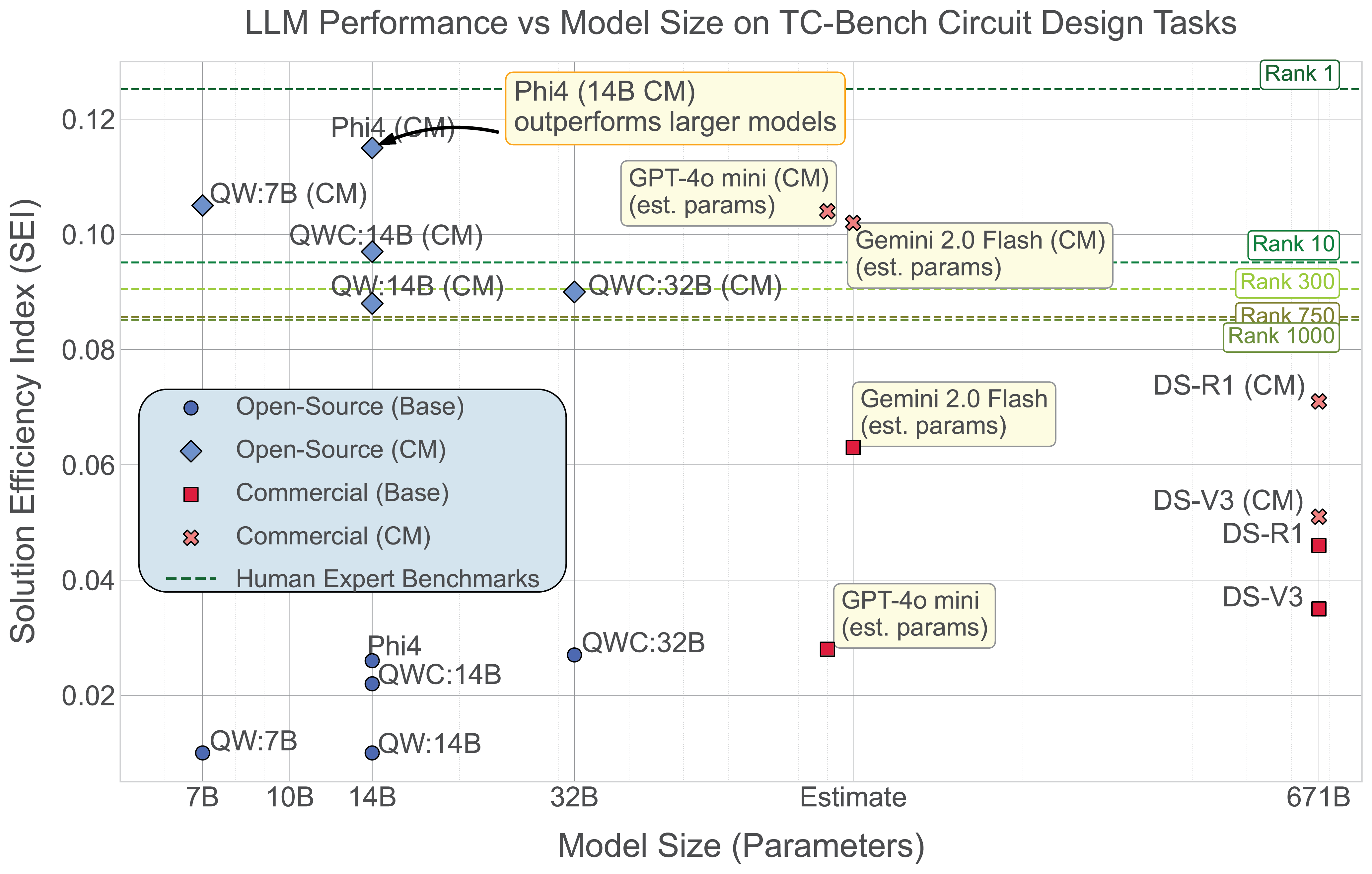}
\caption{SEI vs. Model Size Compared to Human Expert Tiers.}
\label{fig:model_size_sei}
\end{figure}

To assess CircuitMind's real-world significance, we compared its performance against the human expert tiers established in TC-Bench (Fig.~\ref{fig:model_size_sei}).
% --- Placement for Figure 3 (SEI vs Model Size) ---
% ---------------------------------------------------

The results reveal a breakthrough achievement: \textbf{55.6\% of CircuitMind implementations (5 out of 9) reach or exceed top-tier human expert performance}. Specifically:

\begin{itemize}
    \item CM(Phi-4) achieved the highest overall SEI (0.115), firmly placing it in the top-tier human expert range
    \item Four additional implementations reached top-tier performance: CM(Qwen2.5:7B) (0.105), CM(GPT-4o mini) (0.104), CM(Gemini 2.0 Flash) (0.102), and CM(Qwen2.5-Coder:14B) (0.097, near the boundary)
    \item The remaining implementations performed comparably to competent human designers in mid and low tiers
\end{itemize}

Perhaps most significant is that CM(Phi-4), based on a relatively modest 14B-parameter model, outperformed implementations using much larger models. This suggests CircuitMind effectively leverages inherent Boolean reasoning capabilities rather than relying on model scale. This finding is further supported by the DeepSeek models, which achieved high functional correctness but lower efficiency scores (DS-V3: 0.051, DS-R1: 0.071) than several smaller models.

These results indicate that aptitude for Boolean reasoning and compatibility with the framework's collaborative optimization process may be more important than model size or general coding ability for achieving human-competitive efficiency in gate-level circuit design—a direct validation of our architectural approach over scale-based solutions.

\subsection{Component Contribution Analysis}
To isolate the impact of our key technical innovations, we conducted an ablation study focused on the RAG components. We compared three configurations, all maintaining Syntax Locking and Dual-Reward mechanisms:

\begin{itemize}
    \item V0: Baseline CircuitMind with RAG disabled
    \item V1: CircuitMind with only ReviewRAG enabled
    \item V2: Full CircuitMind with both DesignRAG and ReviewRAG enabled
\end{itemize}

The results in Table~\ref{tab:rag_ablation} demonstrate RAG's significant contribution to CircuitMind's performance. Four of seven tested model configurations (DS-R1, 4o-mini, Gemini, Phi-4) showed progressive SEI improvements from V0 to V1 to V2, confirming cumulative benefits from each RAG component. Most dramatically, the full RAG system (V2) improved SEI by 137\% for Gemini (0.043 to 0.102) and 38.6\% for Phi-4 (0.083 to 0.115) compared to the baseline without RAG (V0).
\begin{table}[H] % Using [H] from float package for potentially stricter placement
\centering
\footnotesize
\caption{Ablation Study: Impact of RAG Components.}
\label{tab:rag_ablation}
\resizebox{\columnwidth}{!}{% Add resizebox back if table is too wide
\begin{tabular}{@{}l@{\hspace{2pt}}ccc@{\hspace{4pt}}ccc@{\hspace{4pt}}ccc@{}}
\toprule
\textbf{Model} & \multicolumn{3}{c}{\textbf{Pass@1}} & \multicolumn{3}{c}{\textbf{Pass@5}} & \multicolumn{3}{c}{\textbf{SEI}} \\
\cmidrule(lr){2-4} \cmidrule(lr){5-7} \cmidrule(lr){8-10}
 & \textbf{V0} & \textbf{V1} & \textbf{V2} & \textbf{V0} & \textbf{V1} & \textbf{V2} & \textbf{V0} & \textbf{V1} & \textbf{V2} \\
\midrule
\multicolumn{10}{@{}l}{\textit{Commercial Models}} \\
DS-V3 & .936 & .929 & .964 & .969 & .954 & .990 & .068 & .062 & .051 \\
DS-R1 & .966 & .955 & .968 & .993 & .964 & .997 & .036 & .054 & .071 \\
4o-mini & .750 & .900 & .779 & .909 & .929 & .909 & .052 & .095 & .104 \\
Gemini & .904 & .911 & .888 & .938 & .929 & .944 & .043 & .073 & .102 \\
\midrule
\multicolumn{10}{@{}l}{\textit{Open-Source Models (14B parameters)}} \\
Phi-4 & .643 & .670 & .629 & .836 & .852 & .823 & .083 & .096 & .115 \\
QwenC:14B & .421 & .484 & .500 & .761 & .822 & .805 & .061 & .103 & .097 \\
Qwen:14B & .452 & .468 & .500 & .621 & .655 & .752 & .075 & .062 & .088 \\
\bottomrule
\end{tabular}% End resizebox
}
\vspace{-1em}
\end{table}

While not all models showed strictly monotonic improvement patterns, the overall trend confirms that retrieval mechanisms significantly enhance optimization capabilities, particularly for complex circuits where efficient pattern recognition is crucial.

This study supports the complementary nature of our technical innovations. Syntax Locking establishes the gate-level focus, Dual-Reward directs optimization, and RAG enhances performance by leveraging successful design patterns from the knowledge base. The synergy appears particularly effective for models like Phi-4, where the full V2 system achieves significantly higher efficiency than configurations with partial or no RAG.

\subsection{Case Study: Boolean Optimization}

\begin{table}[h] % Changed to [h] for standard float placement
\centering
\caption{Case Study: Logical Device Implementation Comparison}
\label{tab:case_study}
\begin{tabular}{@{}lcccccc@{}}
\toprule
& \multicolumn{3}{c}{\textbf{Base Model}} & \multicolumn{3}{c}{\textbf{CircuitMind}} \\
\cmidrule(lr){2-4} \cmidrule(lr){5-7}
\textbf{Model} & \textbf{Gates} & \textbf{Delay} & \textbf{SEI} & \textbf{Gates} & \textbf{Delay} & \textbf{SEI} \\
\midrule
GPT-4o mini & 96 & 12 & 0.0093 & 13 & 8 & 0.0476 \\
DeepSeek-V3 & 96 & 12 & 0.0093 & \textbf{8} & \textbf{2} & \textbf{0.1000} \\
Phi-4 & 101 & 16 & 0.0085 & 34 & 8 & 0.0238 \\
\bottomrule
\end{tabular}
\vspace{-1em}
\end{table}

A medium-difficulty logical device implementation case study provides concrete evidence of how CircuitMind overcomes the Boolean Optimization Barrier (Table~\ref{tab:case_study}). Baseline LLMs consistently produced highly redundant circuits with poor efficiency (e.g., 96 gates with SEI ~0.009), exemplifying the limitations described in Section 4.4.

CircuitMind dramatically transformed these implementations:
\begin{itemize}
    \item CM(DeepSeek-V3) reduced the circuit from 96 gates to just 8 gates, achieving a 975\% SEI improvement (0.0093 to 0.1000)
    \item CM(GPT-4o mini) improved efficiency by 412\%, cutting gates from 96 to 13
    \item CM(Phi-4) achieved a 180\% SEI improvement, reducing complexity from 101 gates to 34
\end{itemize}

These results directly demonstrate how CircuitMind guides sophisticated Boolean reasoning, transforming inefficient structures into optimized designs through shared logic identification and Boolean simplification. Most importantly, these improvements were achieved without explicit gate-level training, validating our central thesis that architectural innovation in reasoning distribution can overcome fundamental LLM limitations that scale alone cannot address.

\section{Conclusion}
% \label{sec:conclusion} % Optional label

This research addressed the critical efficiency gap in LLM-based hardware design, identifying the \textit{Boolean Optimization Barrier} and benchmark misalignment as key obstacles. Our proposed multi-agent framework, CircuitMind, integrated with the human-aligned evaluation capabilities of TC-Bench, demonstrated a significant advancement in tackling these challenges.

Experimental results clearly show that our approach substantially improves both functional correctness and, crucially, the physical efficiency (SEI) of AI-generated circuits across diverse LLMs. Achieving human-competitive performance in 55.6\% of implementations, notably with a smaller model like Phi-4 outperforming larger ones, underscores a core finding: intelligent collaborative architecture and targeted optimization strategies, as embodied in CircuitMind, are more pivotal than raw model scale for overcoming gate-level reasoning limitations. TC-Bench proved essential in quantifying this progress against realistic human expert benchmarks.

While current work focuses on gate count and delay, limitations remain regarding power optimization, industrial constraints, and knowledge base scalability. Future efforts should aim to incorporate these broader physical design considerations and enhance the collaborative mechanisms for complex, real-world scenarios.

In conclusion, this work validates that structuring AI systems for collaborative, specialized reasoning, guided by benchmarks reflecting human expertise, offers a powerful pathway to achieving human-level performance in complex engineering optimization tasks like digital circuit design. This paradigm shift emphasizes architectural innovation and holds promise for advancing AI-driven automation in hardware and potentially other physically constrained domains.

% The rest of your document (\bibliographystyle, \bibliography, \end{document}) follows

\bibliographystyle{IEEEtran}

\end{document}